\documentstyle[a4,epsf,psfig,rotate,12pt]{article}
\begin{document}
\epsfclipon
\title{\bf Physics of High-Mass Dimuon Production at the 50-GeV
Proton Synchrotron}
\author{\large J. C. Peng, G. T. Garvey, J. M. Moss\\ 
{\it Physics Division, LANL, Los Alamos, NM 87545, USA}\\ \\
S. Sawada, J. Chiba\\
{\it KEK, 1-1 Oho, Tsukuba, Ibaraki 305-0801, Japan}\\}
\date{\large July 28, 2000}
\maketitle
\bigskip
\begin{abstract}
We discuss the physics interest and the experimental
feasibility for detecting high-mass dimuon pairs using the planned
50-GeV Proton Synchrotron (PS) at the KEK/JHF and JAERI/NSP joint 
accelerator project.
The Drell-Yan measurement of $p+d$ versus $p+p$ at 50 GeV will provide unique 
information on the flavor asymmetry of proton's up and down sea-quark 
distributions in the large-$x$ region. A study of the nuclear dependences
of Drell-Yan cross sections can reveal the modification of antiquark 
distributions in nuclei. Furthermore, the effect of energy loss for fast
partons traversing nuclear medium could also be sensitively measured.
If polarized proton beam becomes available at the 50-GeV PS, 
unique information on the sea-quark polarization could be obtained.
Study of heavy quarkonium production at the 50-GeV PS can set important 
constraints on the mechanism of vector meson productions. Using a
prototype dimuon spectrometer, we have simulated the sensitivities
for a variety of measurements.
\end{abstract}
\newpage
\section{Introduction}
\label{intro}

One of the most active areas of research in nuclear and particle physics 
during the last several decades is the study of quark and gluon distributions 
in the nucleons and nuclei. Several major surprises were discovered in 
Deep-Inelastic Scattering (DIS) experiments which profoundly changed our 
views of the partonic substructure of hadrons. In the early 1980's, the 
famous `EMC' effect found in muon DIS provided the first 
unambiguous evidence that the quark distributions in nuclei are
significantly different from those in free nucleons~\cite{emc83,gee95}. 
More recently, surprising results on the spin and
flavor structures of the nucleons were discovered in DIS experiments. 
The so-called `spin crisis', revealed by the disagreement between the 
prediction of the Ellis-Jaffe sum rule and the polarized DIS 
experiments, has led to extensive theoretical and experimental efforts to 
understand the partonic content of proton's 
spin~\cite{hughes99}. Subsequently, the observation~\cite{nmc91} of the 
violation of the Gottfried sum rule~\cite{gott} in DIS revealed a 
surprisingly large asymmetry between the up and down antiquark distributions 
in the nucleon, shedding new light on the origins of the nucleon sea.\\

The partonic structure of nucleons and nuclei can also be measured with 
hadronic probes. A powerful tool for such studies is the 
Drell-Yan process~\cite{dy71}, 
in which a quark annihilates with an antiquark forming a virtual photon 
which subsequently decays into a lepton pair. The proton-induced
Drell-Yan process is of particular interest, since it can be used to extract 
antiquark distributions of the target nucleon and nuclei. This provides
information complementary to what can be obtained in DIS, which is
sensitive to the sum of the quark and antiquark distributions.\\

The usefulness of the Drell-Yan process as a tool for probing antiquark
distributions has been well demonstrated by a series of Fermilab dimuon
production experiments~\cite{plm99}. In particular, 
the Drell-Yan cross section ratios
for $p+d$ versus $p+p$ led to a direct measurement of the $\bar d/ \bar u$ 
asymmetry as a function of Bjorken-$x$~\cite{hawker98,peng98}. 
Furthermore, the nuclear dependence 
of the Drell-Yan cross sections showed no evidence for antiquark enhancement 
in heavy nuclei~\cite{alde90}, in striking disagreements 
with predictions of some theoretical 
models which were capable of explaining the EMC effect.\\

The 50-GeV Proton Synchrotron (PS) offers a 
unique opportunity to extend existing measurements
of antiquark distributions to much larger values of Bjorken-$x$. Such 
information is crucial for understanding the origins of flavor asymmetry
in the nucleon sea, and for illuminating the nuclear environment effects on
parton distributions. Moreover, Drell-Yan measurements using polarized proton
beam on polarized target at the 50-GeV PS will provide a first determination
of the spin-dependent antiquark distribution at an $x$ region not accessible
in the RHIC-spin program. Indeed, the flavor asymmetry of polarized sea
quark distributions, predicted to be very large by 
certain theoretical models~\cite{dia97,waka99}, can be directly measured.\\

A detailed study of the nuclear dependence of Drell-Yan cross sections at
50 GeV could also lead to a first observation of the coherent partonic
energy-loss effects predicted recently by Baier, Dokshitser, Mueller, Peigne,
Schiff (BDMPS)~\cite{bdmps} and by Zakharov~\cite{zak}. 
These authors studied the radiative energy
loss (through gluon emission) of high energy partons passing through hot 
and cold hadronic matter. The partonic energy-loss 
effect is the QCD analog of the 
Landau-Pomeranchuk-Migdal (LPM) QED effect~\cite{landau,migdal} 
predicted over 40 years ago and confirmed only recently at 
SLAC~\cite{anthony}. A number of surprising effects were
obtained by BDMPS and Zakharov. First, the partonic energy loss in a hot
QCD plasma is predicted to be much larger than in cold matter. This 
suggests that an anomalously large energy loss of 
jets produced in relativistic heavy-ion
collisions could be a signature for Quark-Gluon Plasma formation.
Second, the radiative energy loss is predicted to be proportional
to $L^2$, where $L$ is the path length of hot or cold nuclear matter traversed
by the partons. This curious result is contrary to the conventional wisdom
that energy loss depends linearly on $L$, and it reflects the 
quantum-mechanical interference effect from several contributing diagrams.\\

An attempt to search for partonic energy-loss effects in cold matter 
was made recently via the study of nuclear dependence of Drell-Yan 
cross sections at 800 GeV~\cite{maxim99}. Only an upper limit 
for partonic energy loss was determined. A much more sensitive study
can be made at lower beam energies, where the fractional energy 
loss $\Delta E / E$ will be larger. At the 50-GeV PS, the effect 
is expected to be much enhanced and indeed one could even examine
whether the nuclear effect follows the $L$ or $L^2$ dependence.\\

While logarithmic scaling violation is well established in DIS experiments,
no clear evidence for scaling violation has been seen in Drell-Yan
process. The 50-GeV PS provides an interesting opportunity for unambiguously 
establishing scaling violation in the Drell-Yan process~\cite{peng99}. 
For given values of $x_1$ and $x_2$ (Bjorken-$x$ for the projectile and target 
partons, respectively), scaling-violation is expected to cause roughly 
a factor of two increase in the Drell-Yan cross sections when proton
beam energy is decreased from 800 GeV to 50 GeV. It appears 
quite feasible to establish scaling violation in the Drell-Yan process 
with future data from the 50-GeV PS.\\

Detection of high-mass dileptons at the 50-GeV PS will also allow a study
of $J/\Psi$ and $\Psi^\prime$ production. Existing data on proton-induced
charmonium production are mostly limited to the energy range 
150 GeV $\le E_p \le$ 800 GeV. A comparison of 50 GeV charmonium
production data with existing data will further improve our knowledge
on the production and propagation of charmonium in the nuclear medium.
Many different effects which could affect
the production of charmonium in nuclear medium, such as nuclear shadowing,
partonic energy loss, final-state interaction with comoving gluons or
hadrons, will have different beam energy dependences~\cite{hufner99,vogt00}. 
Therefore, a systematic study of charmonium production in $p-p, p-A$ 
and $A-A$ collisions at 50 GeV would be extremely valuable for 
disentangling various effects. Only after the mechanisms for 
charmonium production are well understood could $J/\Psi$-suppression 
be used confidently as a signature for Quark-Gluon Plasma formation in 
relativistic heavy-ion collisions~\cite{matsui,na50}.\\

Many of the proposed studies for Drell-Yan process at the 50-GeV PS
could also benefit from the $J/\Psi$ and $\Psi^\prime$ data. 
Unlike the Drell-Yan which is an electromagnetic process,
quarkonium production is a strong-interaction process involving
gluon-gluon fusion and quark-antiqaurk fusion. Comparison between the
Drell-Yan and quarkonium production data will further elucidate
various aspects of parton distributions in nucleons and nuclei, and of 
the propagation of partons in nuclei. As an example, charmonium production
with polarized proton beam at the 50-GeV PS might provide interesting 
information on the gluon distributions at large Bjorken-$x$, which 
is essentially unknown.\\

In this paper, we discuss the physics interest and the feasibility
for making precise measurements of 
high-mass dimuons at the 50-GeV PS. In Section 2 the physics 
motivation for various measurements will be discussed in 
some details. A preliminary design study of a dimuon spectrometer for 
the 50-GeV PS will be presented in Section 3. A summary 
is given in Section 4.

\section{Physics Issues}

\subsection{Overview of high-mass dilepton production}

Detection of high-mass dileptons produced in high-energy hadronic interactions
has a long and glorious history. The charm and beauty quarks were discovered
in the 1970's via the dilepton decay modes of 
$J/\Psi$ and $\Upsilon$ resonances.
These quarkonium states are superimposed on a dilepton continuum known as
the Drell-Yan process~\cite{dy71}. The Drell-Yan data has been 
a source of information for the antiquark structure of 
the nucleon~\cite{kenyon}. Furthermore, Drell-Yan production with
pion and kaon beams has yielded the parton distributions of these unstable
particles for the first time. A generalized 
Drell-Yan process was also responsible 
for the discovery of the $W$ and $Z$ gauge bosons in the 1980's.\\

To lowest order, the Drell-Yan process depends on the product of
quark and antiquark distributions in the beam and target as
\begin{equation}
{d^2\sigma\over dx_1 dx_2} = {4\pi\alpha^2\over 9 s~ x_1 x_2}
\sum_a e_a^2[q_a(x_1)
\bar q_a(x_2) + \bar q_a(x_1)q_a(x_2)]. 
\label{eq:dy}
\end{equation}
Here $q_a(x)$ are the quark or antiquark structure functions of 
the two colliding hadrons evaluated at momentum fractions $x_1$ and 
$x_2$. The sum is over quark flavors, and $s$ is the center-of-mass 
energy squared.\\

The kinematics of the virtual photon $-$ longitudinal center-of-mass 
momentum $p_{\|}^{\gamma}$, transverse momentum $p_{T}^{\gamma}$ and 
mass $M_{\gamma}$ $-$ are determined by measuring the two-muon decay 
of the virtual photon.  These quantities determine the momentum 
fractions of the two quarks:
\begin{equation}
x_{F} = p_{\|}^{\gamma} / p_{\|}^{\gamma, max} = x_{1} - x_{2} 
\end{equation}
\begin{equation}
M_{\gamma}^{2} = x_{1} x_{2} s
\end{equation}
where $p_{\|}^{\gamma}$ is the virtual photon center-of-mass longitudinal 
momentum and $p_{\|}^{\gamma, max}$ is the maximum value it can have.\\

\begin{figure}
\centerline{\psfig{file=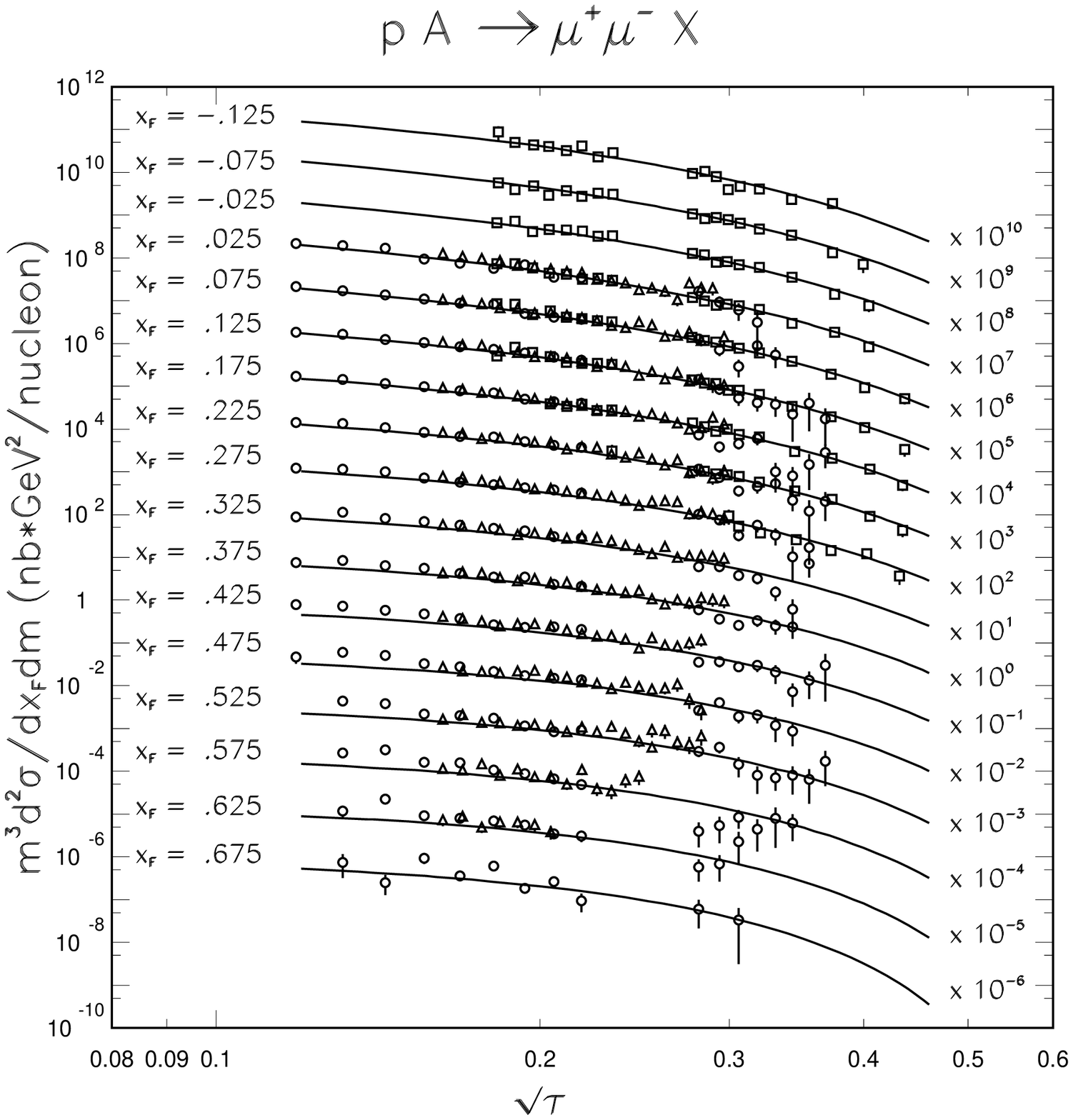,height=6.0in}}
\caption{Proton-induced Drell-Yan production from experiments 
NA3~\cite{na3} (triangles) at 
400 GeV/c, E605~\cite{e605} (squares) at 800 GeV/c, and 
E772~\cite{e772a} (circles) at 800 GeV/c. The lines are absolute (no 
arbitrary normalization factor)
next-to-leading order calculations for 
$p + d$ collisions at 800 GeV/c using the                  
CTEQ4M structure functions~\cite{cteq}.}
\label{figsum}                                                     
\end{figure}

Although the simple parton model 
enjoyed considerable success in explaining many features of the
early data, it was soon realized that QCD corrections to the parton model
were required. 
The inclusion of the NLO diagrams for the Drell-Yan process brings excellent
agreement between the calculations and the data.
As an example, Figure~\ref{figsum} shows the NA3 data~\cite{na3} at 
400 GeV, together with the E605~\cite{e605} and 
E772~\cite{e772a} data at 800 GeV. 
The solid curves in Figure~\ref{figsum} correspond to NLO calculation 
for 800 GeV $p+d$ ($\sqrt s$ = 38.9 GeV) and they describe
the NA3/E605/E772 data well. This shows that the mechanism for Drell-Yan
process is well understood theoretically, and quantitative information on
the parton distributions can be reliably extracted via this process.\\

To gain sensitivity to the antiquark distribution of the target,
one chooses a proton beam and selects the kinematic region of positive $x_F$
and large $x_1$. In this limit the contribution from 
the second term in Eq.~\ref{eq:dy} is small and 
the first term is dominated by the $u(x_{1})$ distribution of the 
proton.  Under these circumstances, the ratio of the cross sections for 
two different targets, $X$ and $Y$, which have $A_{X}$ and $A_{Y}$ nucleons 
is approximately the ratio of the $\bar{u}(x_2)$ distributions:
\begin{equation}
\frac{\frac{1}{A_{X}}\left(\frac{d\sigma^{X}}{dx_{1}dx_{2}}\right)}
     {\frac{1}{A_{Y}}\left(\frac{d\sigma^{Y}}{dx_{1}dx_{2}}\right)}
\approx
\left.\frac{\bar{u}^{X}(x_{2})}{\bar{u}^{Y}(x_{2})}\right|_{x_{1}\gg x_{2}}
\label{eq:csfrac}
\end{equation}
In this relation the cross sections are defined per nucleus but the 
parton distributions are conventionally defined per nucleon.\\

Eq.~\ref{eq:csfrac} demonstrates the power of Drell-Yan experiments 
in determining relative antiquark distributions. This feature was 
explored by recent Fermilab experiments using 800 GeV proton 
beams~\cite{plm99}. The 50-GeV PS provides a unique opportunity for 
extending the Fermilab measurements to larger $x_2$ ($x_2 > 0.25$).
For a given value of $x_1$ and $x_2$, the Drell-Yan cross section
is proportional to $1/s$ (see Eq.~\ref{eq:dy}).
Hence the cross section at 50 GeV is roughly 16 times greater than
that at 800 GeV (The price one pays at lower beam energies is that one 
has limited reach for small $x_2$, which could best be studied at 
higher energies).
Furthermore, to the extent that the radiation dose scales as beam 
power, one can take $\approx$ 16 times higher beam flux
at 50 GeV relative to 800 GeV. The combination of these two effects
could lead to two orders of magnitude improvement in the statistics 
at high $x_2$ over previous Fermilab experiments.\\

\subsection{Scaling violation in Drell-Yan process}

The right-hand side of Eq.~\ref{eq:dy} is only a function of $x_1, x_2$ 
and is independent of the beam energy. This scaling property no longer 
holds when QCD corrections to the Drell-Yan process are taken into account.
While scaling violation is well established in DIS experiments,
it is not confirmed in Drell-Yan experiments at all. 
No convincing evidence for scaling violation is seen~\cite{na10}. 
As discussed in a 
recent review~\cite{plm99}, there are mainly two reasons for this. First, 
unlike the DIS, the Drell-Yan cross section is a convolution of two 
structure functions. For proton-induced Drell-Yan, one often involves 
a beam quark with $x_1 > 0.1$ and a target antiquark with $x_2 < 0.1$. 
Scaling violation implies that the structure
functions rise for $x \leq 0.1$ and drop for $x \geq 0.1$ as $Q^2$
increases. Hence the effects 
of scaling violation are partially cancelled. Second, unlike the DIS, 
the Drell-Yan experiment can only probe relatively large $Q^2$, namely, 
$Q^2 > 16$ GeV$^2$ for a mass cut of 4 GeV. This makes it more difficult 
to observe the logarithmic $Q^2$ variation of the structure functions in 
Drell-Yan experiments.\\

\begin{figure}
\centerline{\psfig{file=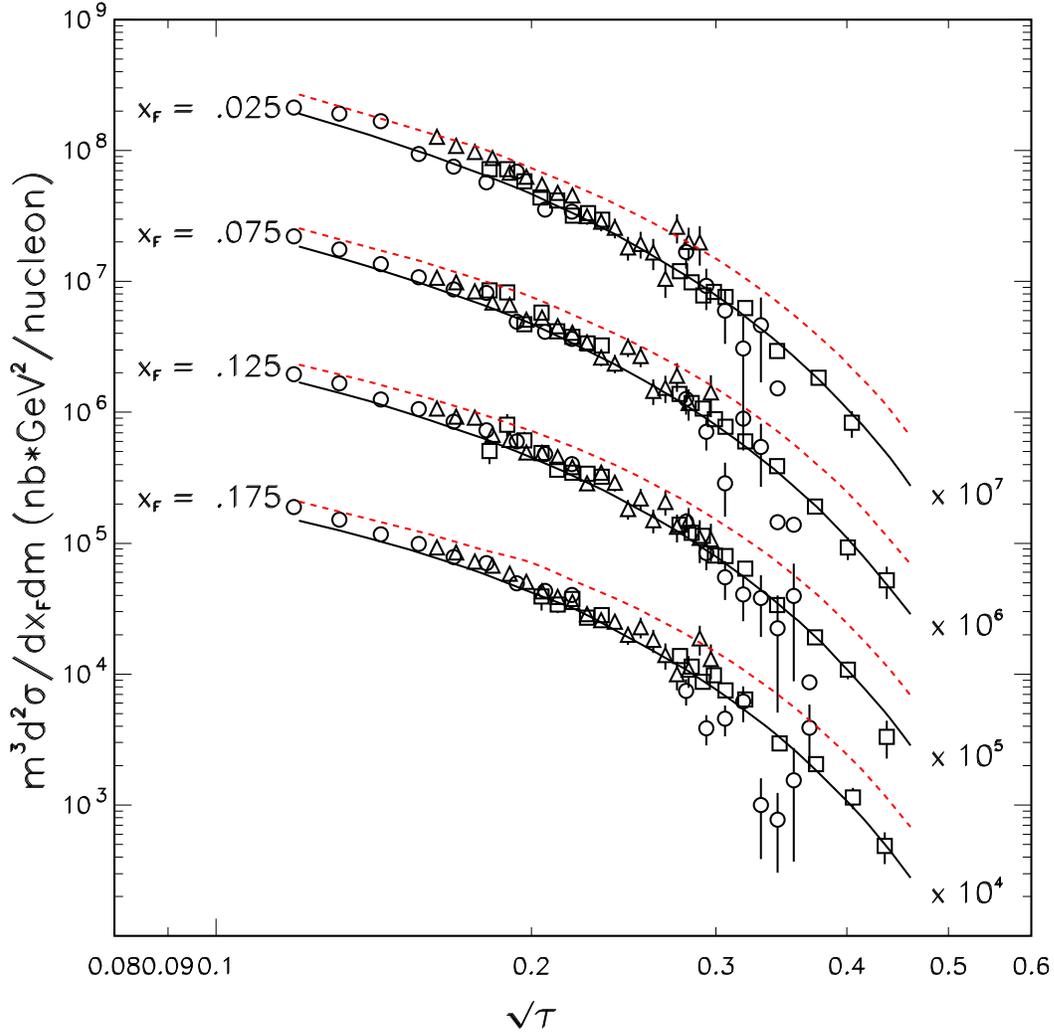,height=6.0in}}
\caption{Comparison of Drell-Yan cross section data with NLO
calculations using MRST~\cite{mrst} structure 
functions. Note that $\tau = x_1 x_2$.
The E772~\cite{e772a}, E605~\cite{e605}, and
NA3~\cite{na3} data points are shown as circles, 
squares, and triangles, respectively.
The solid curves correspond to fixed-target 
$p+d$ collision at 800 GeV, while the
dashed curve is for $p+d$ collision at 50 GeV.}
\label{dyscale}
\end{figure}

The 50-GeV PS provides an interesting opportunity for unambiguously 
establishing scaling violation in the Drell-Yan process. 
Figure~\ref{dyscale} shows the predictions for $p+d$ at 50 GeV.
Scaling violation causes a factor of two increase in the Drell-Yan 
cross sections when the beam energy is decreased from 800 GeV
to 50 GeV. It appears quite feasible to establish scaling violation in
Drell-Yan with future dilepton production experiments at the 50-GeV PS.\\

\subsection{$\bar d/ \bar u$ asymmetry of the proton}

From neutrino-induced DIS experiments, it is known that the strange quark
sea in the nucleon is roughly a factor of two less than the up or down
quark sea~\cite{cdhs}. The lack of SU(3) flavor 
symmetry in the nucleon sea is 
attributed to the much heavier mass of the strange quark. Until recently,
it had been assumed that the distributions of $\bar u$ and $\bar d$
quarks were identical. Although the equality of $\bar u$ and $\bar d$
in the proton is not required by any known symmetry, this is a plausible
assumption for sea quarks generated by gluon splitting. Because 
the masses of the 
up and down quarks are small compared to the confinement scale, nearly equal 
numbers of up and down sea quarks should result.\\

The assumption of $\bar u(x) = \bar d(x)$ can be tested by measurements
of the Gottfried integral~\cite{gott}, defined as
\begin{equation}
I_G = \int_0^1 \left[F^p_2 (x,Q^2) - F^n_2 (x,Q^2)\right]/x~ dx =
{1\over 3}+{2\over 3}\int_0^1 \left[\bar u_p(x)-\bar d_p(x)\right]dx,
\label{eq:3.1}
\end{equation}
where $F^p_2$ and $F^n_2$ are the proton and neutron structure
functions measured in DIS experiments. 
Under the assumption of a symmetric sea, $\bar u$ = $\bar d$,
the Gottfried Sum Rule (GSR)~\cite{gott}, $I_G
= 1/3$, is obtained. 
The most accurate test of the GSR was reported in 1991 by the New Muon 
Collaboration (NMC)~\cite{nmc91}, which measured $F^p_2$ and $F^n_2$ over the 
region $0.004 \le x \le 0.8$. They determined the Gottfried integral to be 
$ 0.235\pm 0.026$, significantly below 1/3. This surprising result has
generated much interest, and it strongly suggested that the assumption
$\bar u = \bar d$ should be abandoned.  Specifically, the NMC result implies
\begin{equation}
\int_0^1 \left[\bar d(x) - \bar u(x)\right] dx = 0.148 \pm 0.039.
\label{eq:3.2}
\end{equation}
Eq.~\ref{eq:3.2} shows that only the integral of $\bar d -\bar u$ was deduced
from the DIS measurements. The $x$ dependence of $\bar d - \bar u$ remained
unspecified.\\

The proton-induced Drell-Yan process provides an
independent means to probe the flavor asymmetry of the nucleon sea~\cite{es}.
An important advantage of the Drell-Yan process is that the $x$ dependence of 
$\bar d / \bar u$ can be determined.
The Fermilab E772 collaboration~\cite{plm}
compared the Drell-Yan yields from 
isoscalar targets with that from a neutron-rich 
(tungsten) target, and constraints on the nonequality of $\bar u$ and 
$\bar d$ in the range $0.04 \leq x \leq 0.27$ were set. 
More recently, the CERN experiment NA51~\cite{na51}
carried out a comparison of the Drell-Yan muon pair yield from hydrogen and
deuterium using a 450 GeV/c proton beam. They found that
$\bar u / \bar d = 0.51 \pm 0.04 \pm 0.05$ at $\langle x \rangle =0.18$, a 
surprisingly large difference between the $\bar u$ and $\bar d$.\\

\begin{figure}
\center
\hspace*{0.5in}
\psfig{figure=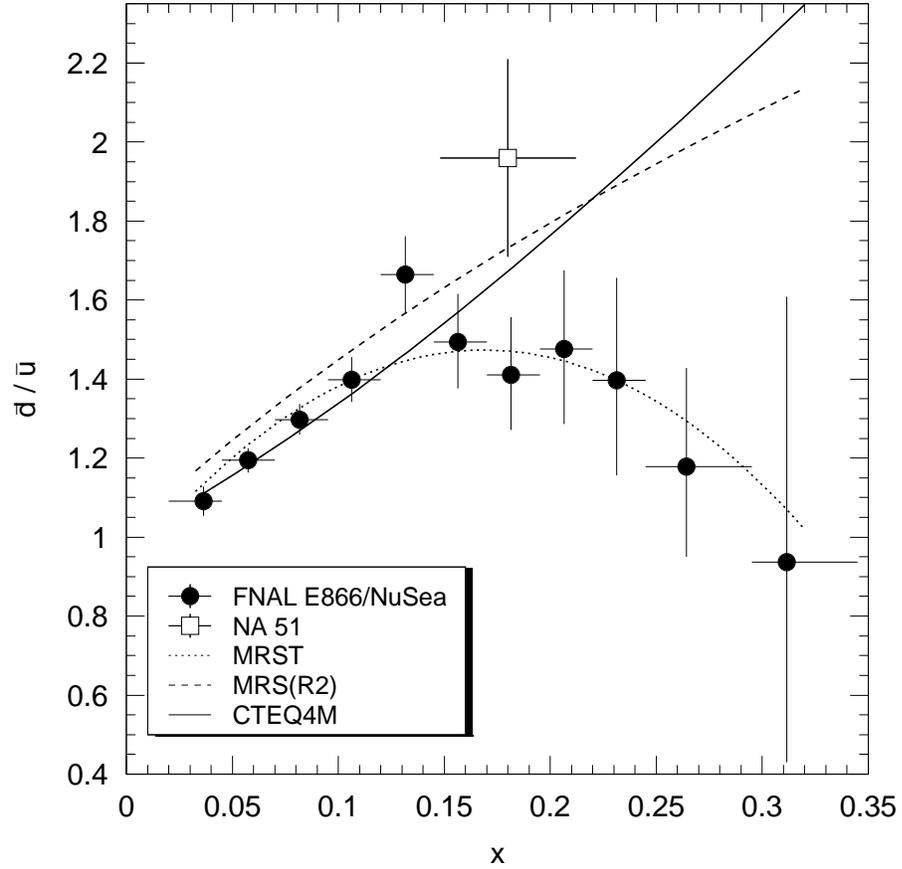,height=4.68in}
  \caption{The ratio of $\bar{d}/\bar{u}$ in the proton as a function
  of $x$ extracted from the Fermilab E866~\cite{hawker98} cross section ratio. The
  curves are from various parton distributions.  
  Also shown is the result
  from NA51~\cite{na51}, plotted as an open square.}
\label{fig:3.2}
\end{figure}

A Drell-Yan experiment (E866), aiming at higher statistical accuracy and wider 
kinematic coverage than NA51, was recently completed~\cite{hawker98,peng98}
at Fermilab. This experiment also measured the Drell-Yan muon pairs from 
800-GeV/c protons interacting with liquid deuterium and hydrogen targets.
Eq.~\ref{eq:csfrac} shows that the Drell-Yan cross section ratio
at large $x_F$ is approximately given as 
\begin{equation}
{\sigma_{DY}(p+d)\over 2\sigma_{DY}(p+p)} \approx
{1\over 2} \left(1+{\bar d(x_2)\over \bar u(x_2)}\right).
\label{eq:3.3}
\end{equation}
Values for $\bar d/ \bar u$ were extracted by the E866 collaboration at
$Q^2 = 54$ GeV$^2$/c$^2$ over the region $0.02 < x < 0.345$.
These are shown in Figure~\ref{fig:3.2} along 
with the NA51 measurement. For $x < 0.15$, $\bar d/\bar u$ increases 
linearly with $x$ and is in good agreement with the CTEQ4M~\cite{cteq} 
and MRS(R2)~\cite{mrs} parameterizations. However, a distinct feature 
of the data, not seen in either parameterization, is the rapid decrease 
toward unity of $\bar{d}/\bar{u}$ beyond $x=0.2$.\\

\begin{figure}
\center
\hspace*{0.5in}
\psfig{figure=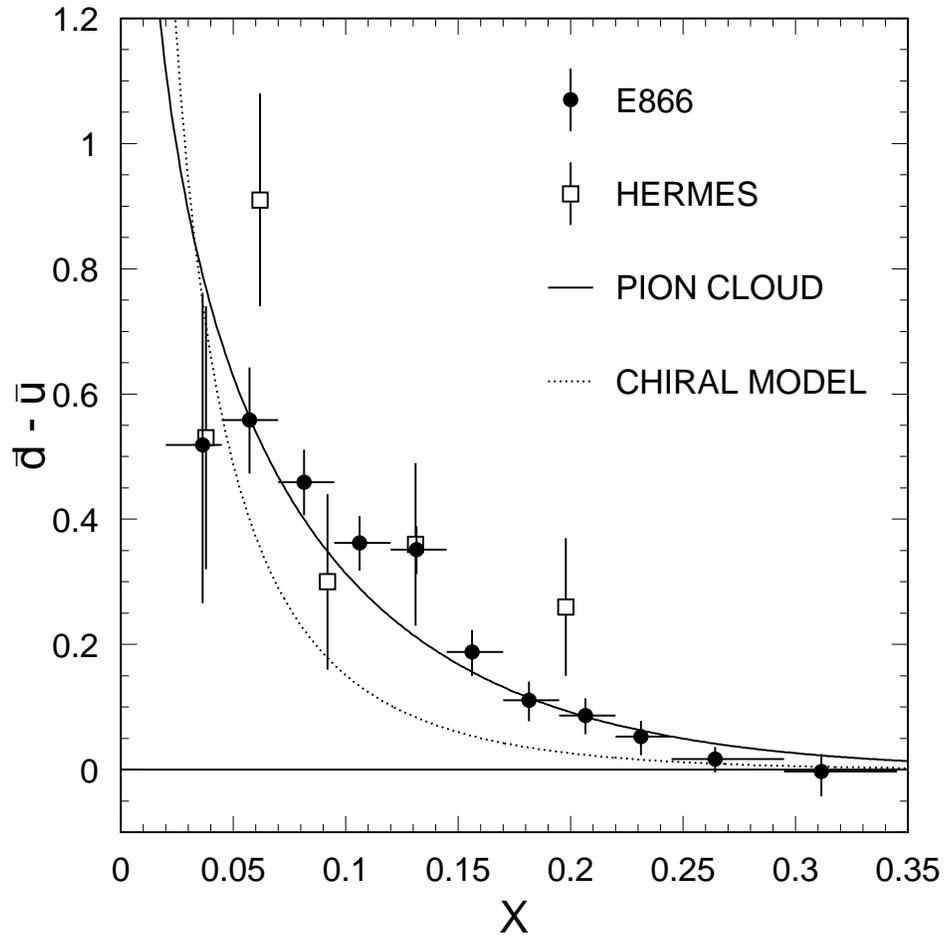,height=5.4in}
\caption{Comparison of the E866~\cite{hawker98} $\bar d - \bar u$ results at $Q^2$ =
54 GeV$^2$/c$^2$ with the predictions of pion-cloud and chiral models 
as described in the text. The data from HERMES~\cite{hermes} are also shown.}
\label{fig:3.3}
\end{figure}

The $\bar d / \bar u$ ratio, along with the CTEQ4M values for 
$\bar d + \bar u$, was used to obtain $\bar d - \bar u$ 
(Figure~\ref{fig:3.3}). Being a flavor nonsinglet quantity,
$\bar d(x) - \bar u(x)$ is decoupled from gluon distribution.
Since perturbative processes have negligible contribution
to $\bar d / \bar u$ asymmetry, $\bar d(x) - \bar u(x)$ essentially
isolates the contribution from non-perturbative effects.
From the results shown in Figure~\ref{fig:3.3}, one can
obtain an independent determination~\cite{peng98} of the integral of 
Eq.~\ref{eq:3.2}.
E866 finds $0.100 \pm 0.007
\pm 0.017$, consistent with, but roughly $2/3$ of the value deduced by NMC.\\

As early as 1983, Thomas~\cite{thomas} pointed out that the virtual pions
that dress the proton will lead to an enhancement of $\bar d$ relative
to $\bar u$ via the (nonperturbative) 
``Sullivan process.''
Sullivan~\cite{sullivan} previously showed that in DIS 
virtual mesons scale in the Bjorken limit and contribute to the
nucleon structure function.  Following the publication of the NMC
result, many papers treated virtual mesons as the origin of the 
$\bar d/\bar u$ asymmetry
(see~\cite{kumano0} for a recent review).
Here the $\pi^+(\bar d u)$ cloud, dominant in the 
process $p\rightarrow\pi^+ n$, leads to an excess of $\bar d$ sea.\\

A different approach for including the effects of virtual mesons has
been presented by Eichten et al.~\cite{ehq} and
further investigated by other authors~\cite{cheng1,szczurek2}. In 
chiral perturbation theory, the relevant degrees of
freedom are constituent quarks, gluons, and Goldstone bosons. In
this model, a portion of the sea comes from the couplings of Goldstone
bosons to the constituent quarks, such as $u \to d \pi^+$ and $d \to u
\pi^-$. The excess of $\bar d$ over $\bar u$ is then simply due to the
additional valence $u$ quark in the proton.\\

The $x$ dependence of $\bar d - \bar u$ and $\bar d / \bar u$ obtained
by E866 provides important constraints for theoretical models.
Figure~\ref{fig:3.3} compares $\bar d(x) - \bar u(x)$ from E866
with a virtual-pion model calculation, following the procedure detailed
by Kumano~\cite{kumano}. A dipole form, with $\Lambda = 1.0$ GeV for 
the $\pi N N$ form factor and $\Lambda = 0.8$ GeV for 
the $\pi N \Delta$ form factor,
was used. $\Lambda$ is the cutoff parameter for the pion form factor.
Figure~\ref{fig:3.3} also shows the predicted
$\bar d - \bar u$ from the chiral model~\cite{szczurek2}. 
The chiral model places 
more strength at low $x$ than does the virtual-pion
model. This difference reflects the fact that the pions are softer in 
the chiral model, since they are coupled to constituent quarks that 
carry only a fraction of the nucleon momentum. The $x$ dependence of 
the E866 data favors the virtual-pion model over the chiral model, 
suggesting that correlations between the chiral constituents should be 
taken into account.\\

Recently, the flavor asymmetry of the nucelon sea was computed 
in the large-$N_c$ limit, where the nucleon is described as a 
soliton of an effective chiral theory~\cite{waka99, poby99}.
In this chiral quark-soliton model, the flavor non-singlet 
distribution, $\bar d(x) - \bar u(x)$, appears in the next-to-leading
order of the $1/N_c$ expansion~\cite{dia97}. The E866 
$\bar d(x) - \bar u(x)$ data were shown to be well described by 
this model~\cite{poby99}.\\

Instantons have been known as theoretical constructs since the
seventies~\cite{bel75,hooft76,shu98}. 
They represent non-perturbative fluctuations of the gauge fields
that induce transitions between degenerate ground states of different
topology. In the case of QCD, the collision between a quark and an
instanton flips the helicity of the quark while creating a $q \bar q$
pair of different flavor. Thus, interaction between a $u$ quark and 
an instanton results in a $u$ quark of opposite helicity and 
either a $d \bar d$ or $s \bar s$ pair. 
Such a model has the possibility of accounting for both the flavor
asymmetry and the ``spin crisis"~\cite{forte89,forte91}. 
However, the prediction~\cite{inst93} at large 
$x$, $\bar d(x) / \bar u(x) \to 4$, 
is grossly violated by experiment (see Figure~\ref{fig:3.2}). 
Thus, it appears that while instantons have the
possibility for accounting for flavor and spin anomalies, the approach is not
yet sufficiently developed for a direct comparison.\\

The interplay between the perturbative and non-perturbative components of
the nucleon sea remains to be better determined. Since the perturbative
process gives a symmetric $\bar d/ \bar u$ while a non-perturbative 
process is needed to generate an asymmetric $\bar d/ \bar u$ sea, the relative
importance of these two components is directly reflected in the $\bar d/ \bar u$
ratios. Thus, it would be very important to extend the Drell-Yan
measurements to kinematic regimes beyond the current limits.\\

\begin{figure}
\center
\hspace*{0.5in}
\psfig{figure=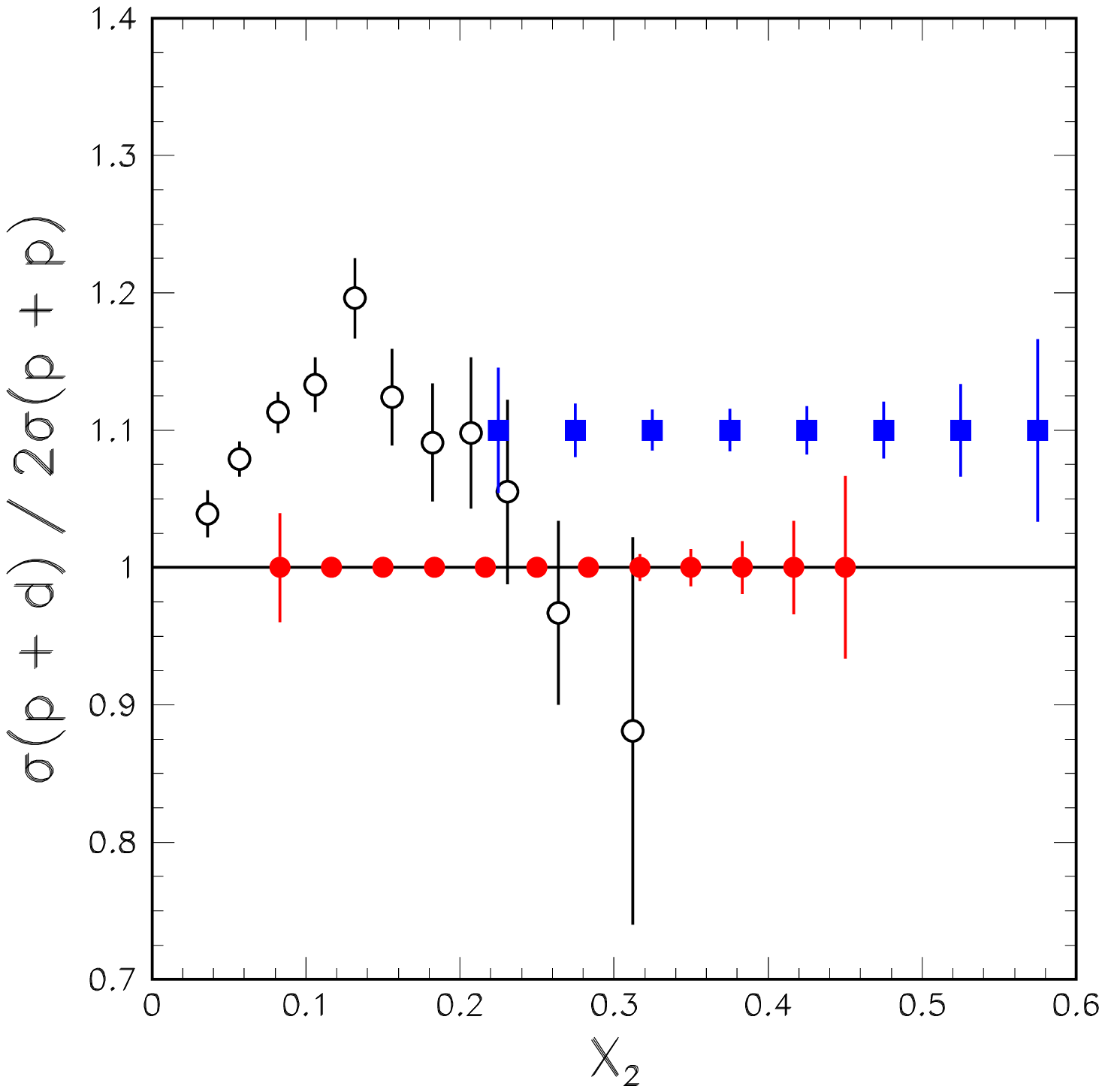,height=5.7in}
\caption{$(p+d) / (p+p)$ Drell-Yan ratios from E866 (open circles) are compared
with the expected sensitivites at the 120 GeV Main Injector (solid circles)
and the 50-GeV PS (solid squares).}
\label{fig:rate1}
\end{figure}

The 50-GeV PS presents an excellent opportunity for 
extending the $\bar d/ \bar u$ measurement to larger $x$ ($x > 0.25$).
As mentioned earlier, for given values of $x_1$ and $x_2$ the Drell-Yan
cross section is proportional to $1/s$, hence the Drell-Yan cross section 
at 50 GeV is roughly 16 times greater than at 800 GeV. 
Figure~\ref{fig:rate1} shows the expected statistical 
accuracy for $\sigma (p+d)/ 2 \sigma (p+p)$ at the 50-GeV PS (see Section 3)
compared 
with the data from E866 and a proposed measurement~\cite{p906} using 
the 120 GeV proton beam at the Fermilab Main-Injector. 
A definitive measurement of
the $\bar d/ \bar u$ over the region $0.25 < x < 0.7$ could indeed be
obtained at the 50-GeV PS.

\subsection{Polarized Drell-Yan at the 50-GeV PS}

Despite extensive work on polarized DIS, the helicity distributions
of $\bar u$ and $\bar d$ sea quarks are still poorly known.
Both the SMC~\cite{smc98} and the HERMES~\cite{hermes99} 
experiments attempted to extract
the sea-quark polarizations via semi-inclusive polarized DIS measurements,
and the results indicate small sea-quark polarization consistent with
zero. However, as pointed out in Ref.~\cite{dress99}, large uncertainties
are associated with certain assumptions made in the extraction.\\

\begin{figure}
\center
\hspace*{0.5in}
\psfig{figure=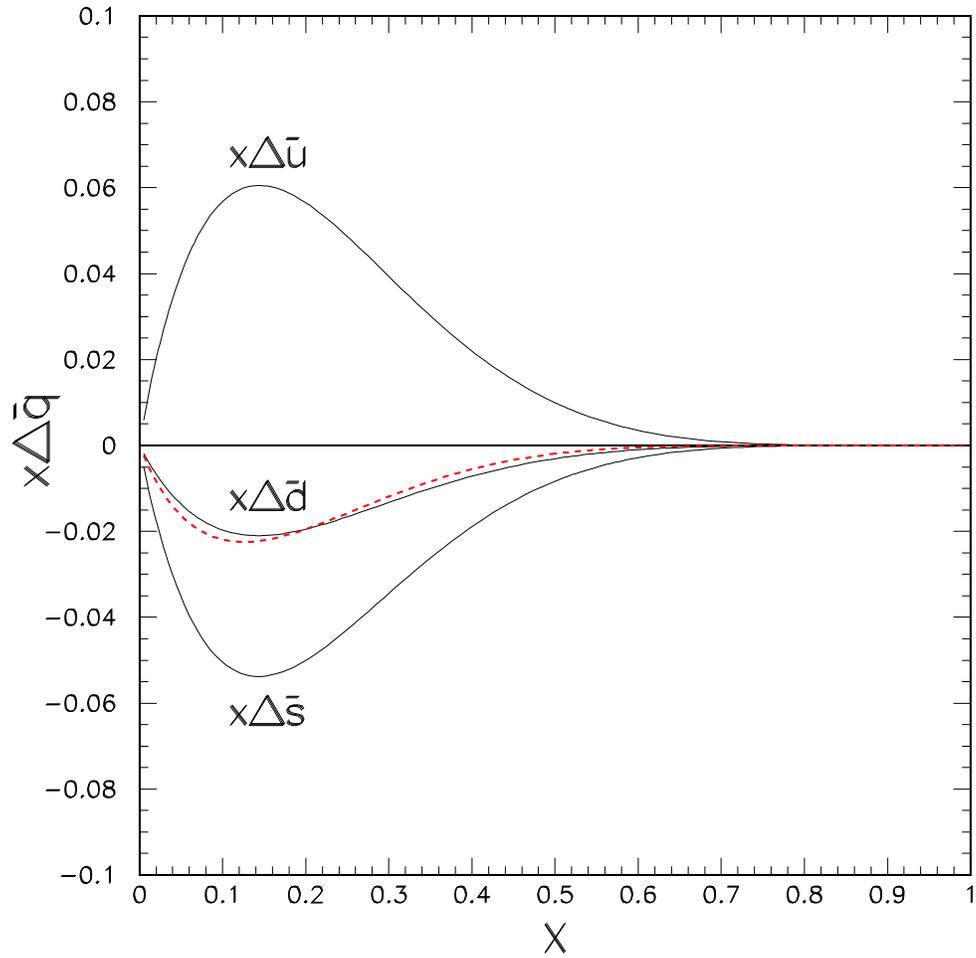,height=5.7in}
\caption{Sea quark polarizations at $Q^2$=0.36 GeV$^2$ predicted by 
chiral-quark soliton model are shown as 
solid curves. The dashed curve corresponds
to the parametrization of $x\Delta \bar q$ at $Q^2 = 0.4$GeV$^2$ by 
GRSV, where $\Delta \bar q = \Delta \bar u = \Delta \bar d = \Delta \bar s$.}
\label{seapol1}
\end{figure}

A direct measurement of sea-quark's polarization is clearly very
important for understanding the flavor decomposition of proton's spin.
Different theoretical models make drastically different
predictions. In particular, the meson-cloud models, which successfully
describe the unpolarized $\bar d$/$\bar u$ asymmetry, predict negligible
amount of sea-quark polarization~\cite{fries98,bores99}. 
Several current parametrizations~\cite{gs96,grsv96} of
polarized parton distributions also assume very small polarization 
for sea quarks. The chiral-quark soliton model, on the other hand, 
predicts substantial sea-quark polarization~\cite{dia97,dress99}. 
Figure~\ref{seapol1}
shows $x\Delta \bar u(x)$, $x\Delta \bar d(x)$, and $x\Delta \bar s(x)$ 
at $Q_0^2 = 0.36$ GeV$^2$ from a recent prediction of chiral-quark
soliton model~\cite{goeke00}. Also shown in Figure~\ref{seapol1}
are the GRSV parametrizations~\cite{grsv96} from a global 
fit to polarized DIS data.\\

\begin{figure}
\center
\hspace*{0.5in}
\psfig{figure=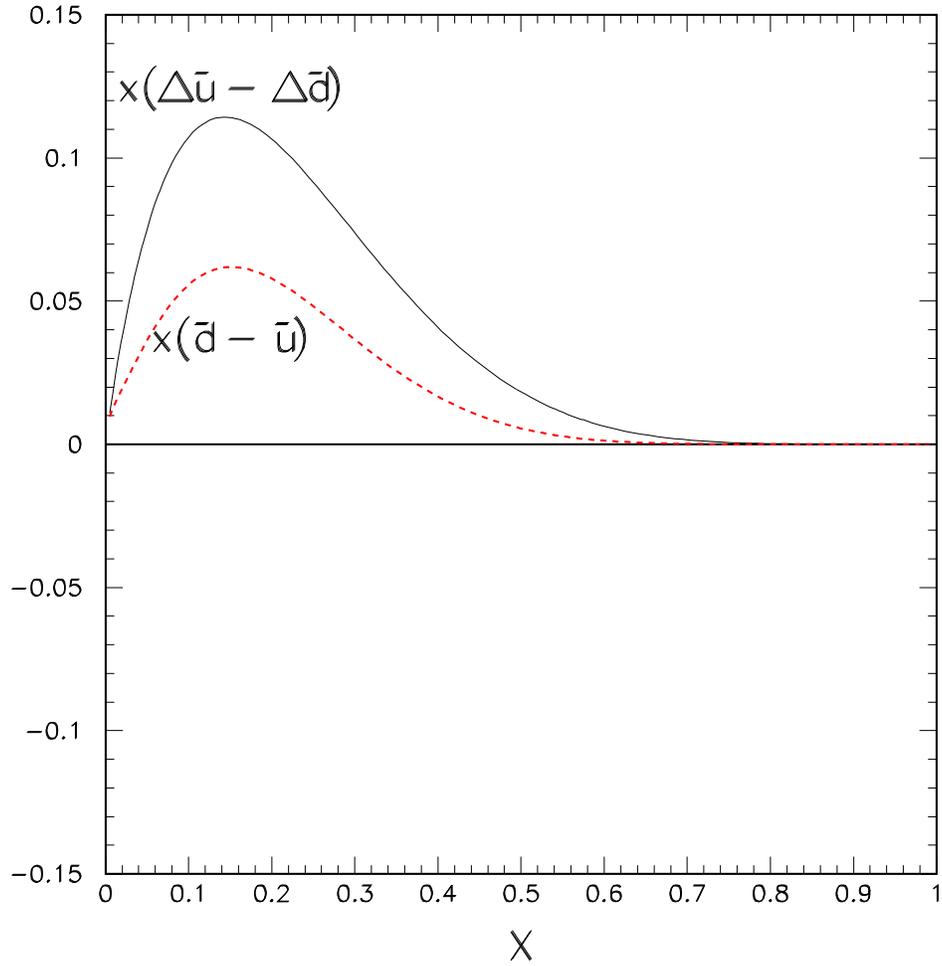,height=5.7in}
\caption{$x(\Delta \bar u - \Delta \bar d)$ at $Q^2 = 0.36$ GeV$^2$ predicted
by the chiral-quark soliton model is shown as the solid curve. The GRV94 LO
parametrization of $x(\bar d - \bar u)$ at $Q^2 = 0.4$ GeV$^2$ is shown as the 
dashed curve.}
\label{seapol2}
\end{figure}

A very striking prediction of the chiral-quark model is the large
flavor asymmetry of polarized sea-quark polarization. In fact, this model
predicts a significantly larger values for $\Delta \bar u - \Delta \bar d$
than for $\bar d - \bar u$. This is shown in Figure~\ref{seapol2}, where
$x(\Delta \bar u - \Delta \bar d)$ from the chiral-quark 
soliton model~\cite{goeke00} is compared with the $x(\bar d - \bar u)$
parametrization from GRV94~\cite{grv94}.\\

\begin{figure}
\center
\hspace*{0.5in}
\psfig{figure=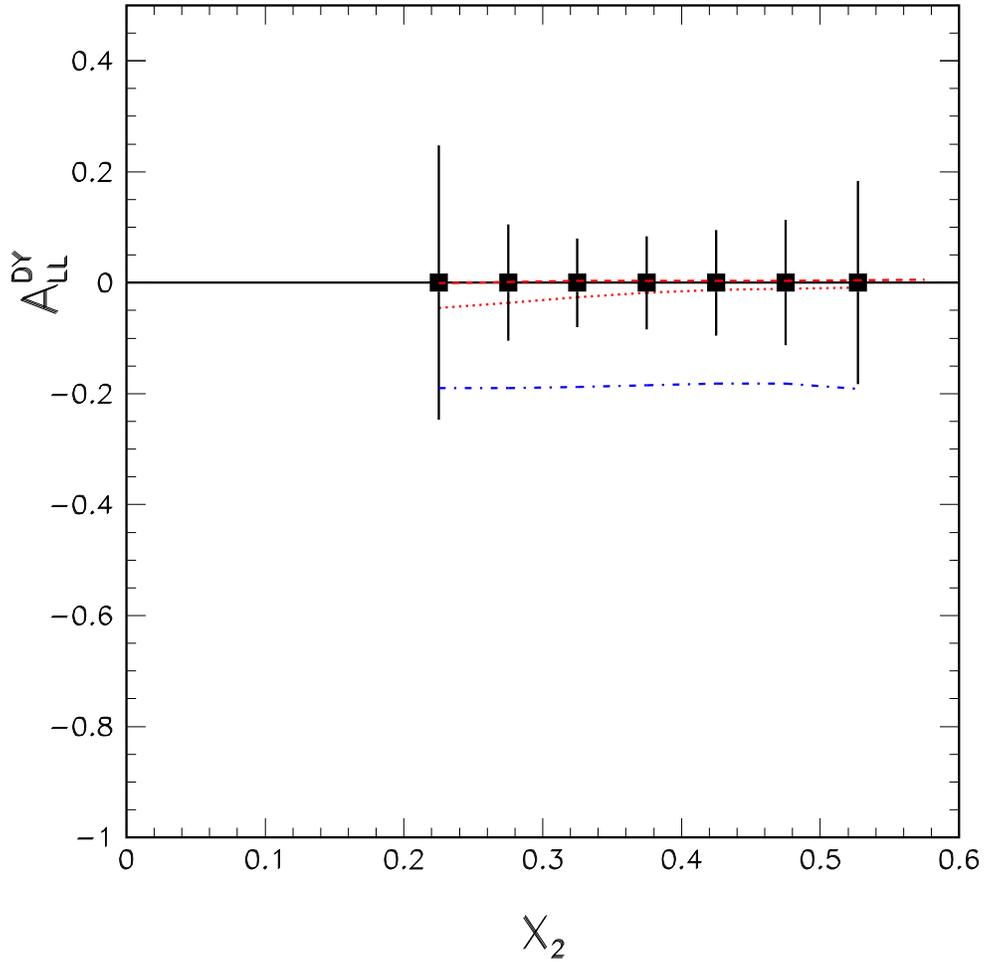,height=5.7in}
\caption{Expected statistical accuracy for measuring the
double-helicity asymmetry $A_{LL}^{DY}$ in polarized $p + p$
Drell-Yan at the 50-GeV PS for a 120-day run. The dashed, dotted,
and dash-dotted curves correspond to calculations using polarized
PDF parametrization of G-S (set A, set C) and GRSV, respectively.}
\label{fig:rate5}
\end{figure}

Polarized proton beam at the 50-GeV PS would offer an exciting
opportunity for probing sea-quark polarizations. 
The longitudinal spin asymmetry in the DY process is, in leading order, 
given by~\cite{close},
\begin{eqnarray}
A_{LL}^{DY}(x_1,x_2)={\sum_a e_a^2[\Delta q_a(x_1) \Delta \bar
q_a(x_2)+\Delta\bar q_a(x_1) \Delta q_a(x_2)]\over\sum_a e_a^2[q_a(x_1)
\bar q_a(x_2)+\bar q_a(x_1)q_a(x_2)] }, \label{eq:ALL0} 
\end{eqnarray}
with $\Delta q_a\equiv q_a^+-q_a^-$.  The superscripts refer to parton
spin projections parallel ($+$) or antiparallel ($-$) to the proton's
spin projection. 
We have simulated the performance of the proposed high-mass dimuon 
spectrometer for measuring polarized antiquark distribution. 
Figure~\ref{fig:rate5} shows the $x_2$ dependence of $A^{DY}_{LL}$,
integrated over the spectrometer acceptance, for polarized
sea-quark parametrizations including Gehrmann-Stirling (G-S) 
sets A and C~\cite{gs96} and GRSV Leading-Order set~\cite{grsv96}.
Very small values for $A_{LL}^{DY}$ are predicted for the G-S
parametrization, while the GRSV parametrization gives 
$A_{LL}^{DY} \approx -0.2$. 
The chiral-quark soliton model gives large positive $A_{LL}^{DY}$.
In fact, the positivity requirement, namely,
$-1 < \Delta \bar u(x) / \bar u(x) < 1$, is not always satisfied at the region
$x > 0.2$ for the particular parametrization given by Ref.~\cite{goeke00}.\\

We have calculated the
expected statistical sensitivities for a 120-day $\vec p + \vec p$
measurement, assuming $75 \%$ polarization for a $5 \times 10^{11}$
per spill polarized proton beam. We also assume a polarized solid
NH$_3$ target similar to the one used by the SMC~\cite{adams99} which 
achieves a hydrogen polarization of $75 \%$ 
and a dilution factor of 0.15. The target length is chosen to give the
same gm/cm$^2$ as for the liquid deuterium target.
Figure~\ref{fig:rate5} shows that the statistical accuracy
of such a measurement can well test the predictions of various
model (note that the chiral-quark soliton model predicts a large positive
$A_{LL}^{DY}$ not shown in this figure). A comparison
of $\vec p + \vec p$ with $\vec p + \vec d$ will further determine
$\Delta \bar d$, which provides a direct test of the chiral-quark
soliton model's prediction of large $\Delta \bar u - \Delta \bar d$.\\

In the polarized Drell-Yan process one may 
also measure a new structure function,
called transversity, which is a correlation between quark 
momentum and its perpendicular spin component~\cite{ralston}. 
The transversity is not measurable in inclusive DIS~\cite{jaffe}.
It is measurable, in 
principle, in collisions of polarized protons whose spins are aligned 
perpendicular to the plane of dilepton detection~\cite{kumano00}. 
A non-zero transverse 
spin correlation in the Drell-Yan
process would clearly require both quark and antiquark transversities to be 
non-zero. Polarized proton beam at the 50-GeV PS could provide
unique information on the transversities at large $x$.

\subsection{Nuclear effects of Drell-Yan}

\begin{figure}
\center
\hspace*{0.5in}
\psfig{figure=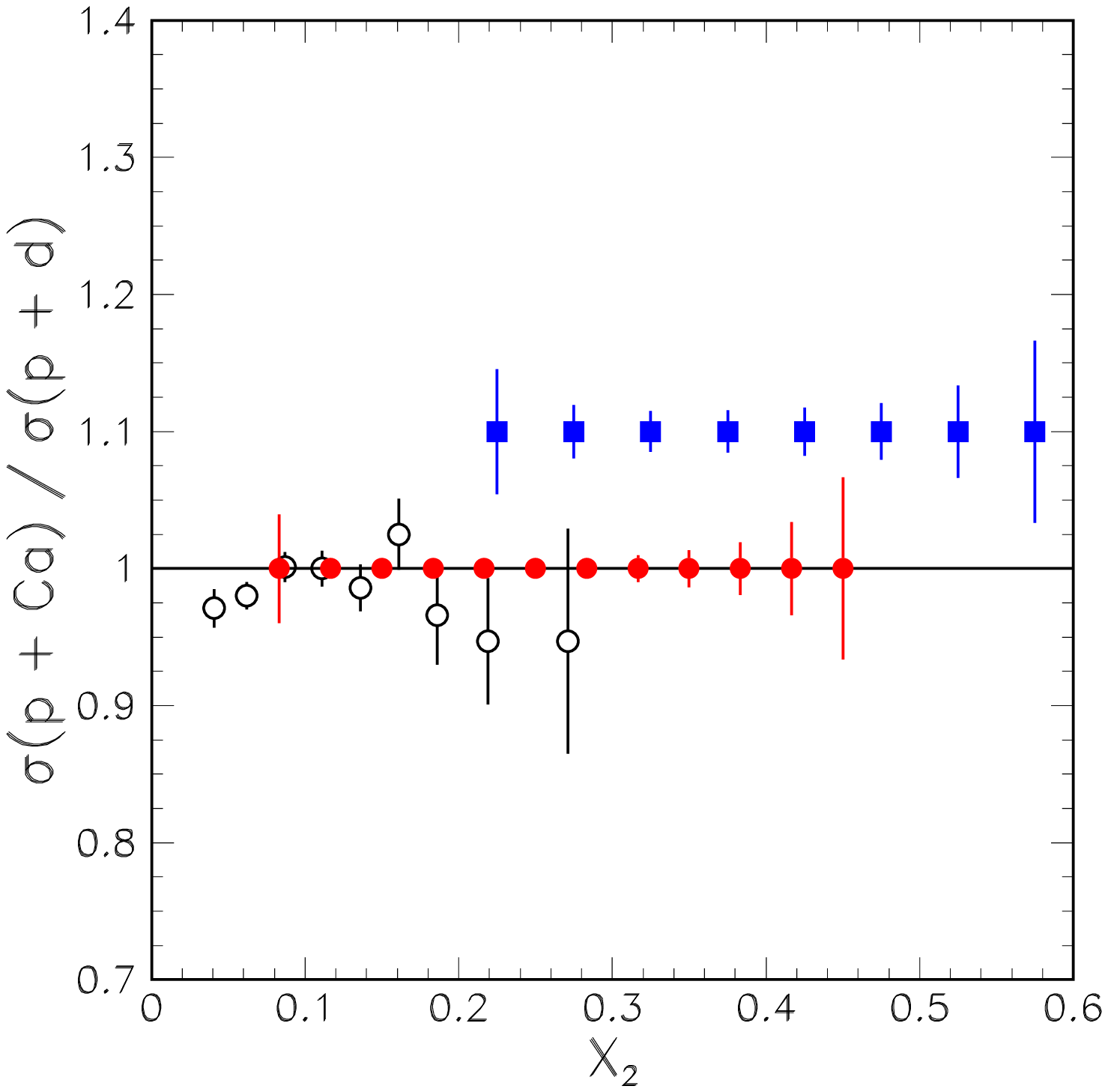,height=6.0in}
\caption{$(p+Ca) / (p+d)$ Drell-Yan ratios from E772 (open circles) are compared
with the expected sensitivites at the 120 GeV Main Injector (solid circles)
and at the 50 GeV PS (solid squares).}
\label{fig:rate6}
\end{figure}

Following the discovery of the EMC effect, it was 
suggested~\cite{pion,pion1,pion2} 
that this effect is caused by the excess of virtual pions in nuclei, which
significantly modify the nuclear parton distributions. A direct 
consequence of the ``pion-excess" model is the nuclear enhancement 
of antiquark sea, which can be probed via Drell-Yan experiment~\cite{miller}. 
However, the subsequent Fermilab E772 experiment~\cite{alde90}
found no evidence for such enhancement (see Figure~\ref{fig:rate6}). The lack
of an antiquark enhancement in nuclei suggests that there are no more pions
surrounding an average nucleon in a heavy nucleus than there are in a
weakly bound system, deuterium. This contradicts conventional wisdom and is
also at odds with sophisticated calculations 
using realistic nuclear force~\cite{pand}.
Unfortunately, the error bars for the E772 data in the region
$x > 0.15$ become quite large, due entirely to limited statistics.
Furthermore, at $x < 0.1$ the on-set of the shadowing effect makes the isolation
of possible pion-excess effect somewhat uncertain.\\

At the 50-GeV PS, one can measure the nuclear effect over the large $x$ region
($x > 0.15$) with high accuracy. This is illustrated in Figure~\ref{fig:rate6},
where the expected statistical errors for a 60-day measurement of $p+Ca$
and $p+d$ using the proposed spectrometer (see Section 3) are shown.
One advantage of the 50-GeV measurement is that shadowing effect is no longer
important at large $x$. The precise measurement at $x$ larger than E772
could access would provide extremely valuable new information on the nuclear
dependence of antiquark distributions. The anticipated sensitivity will be
sufficient to observe the reduction in the nuclear sea distributions
predicted in the $Q^2$ rescaling models~\cite{clj}. 
The pion-excess model, on the
other hand, predicts a strong nuclear enhancement of Drell-Yan
cross sections in this $x$ region.

\subsection{Partonic energy loss in nuclei}

The subject of energy loss of fast partons propagating through hadronic matters
has attracted considerable interest recently~\cite{baier00}. 
The nuclear dependence
of the Drell-Yan process provides a particularly clean way to measure 
the energy loss of incident quarks in a cold nuclear medium.
Partonic energy loss would lead to a degradation of the quark momentum 
prior to annihilation, resulting in a less energetic muon pair.
Therefore, one expects the Drell-Yan cross sections for heavier nuclear
targets to drop more rapidly at large $x_1$ (or $x_F$).\\

Data from E772 at 800 GeV/c were analysed by 
Gavin and Milana~\cite{gavin92} to deduce
the initial-state quark energy loss. They ignored the shadowing effect
and assumed the following expression for the average change in
the momentum fraction:
\begin{equation}
\Delta x_1 = - \kappa_1 x_1 A^{1/3}.
\label{eq:gavin}
\end{equation}
A surprisingly large fractional energy loss 
($\approx 0.4\%/fm$) was obtained. This result was questioned by 
Brodsky and Hoyer~\cite{brodsky93}, who argued that the time scale for gluon bremsstrahlung
need to be taken into account. Moreover, as pointed out in 
Ref.~\cite{plm99}, it is important to account for 
the shadowing effect before a reliable value of partonic
energy loss can be extracted. Using an analogy to the photon
bremsstrahlung process, Brodsky and Hoyer suggested an alternative
expression:
\begin{equation}
\Delta x_1 \approx - {\kappa_2 \over s} A^{1/3},
\label{eq:brodsky}
\end{equation}
where $s$ is the square of the nucleon-nucleon center-of-mass energy.
Note that Eq.~\ref{eq:gavin} implies a linear dependence of
the energy loss on the partonic energy, while Eq.~\ref{eq:brodsky}
assumes a constant energy loss independent of the partonic energy
(note that $\Delta E$ is proportional to $\Delta x_1 s$). Based on
uncertainty principle, Brodsky and Hoyer concluded that energetic
partons should loose $\le 0.5$ GeV/fm in nuclei. More recently,
Baier et al.~\cite{bdmps} and Zakharov~\cite{zak} predicted
\begin{equation}                                                  
\Delta x_1 \approx - {\kappa_3 \over s} A^{2/3}.
\label{eq:baier}
\end{equation}
These authors obtained the nonintuitive result that the total energy
loss is proportional to the square of the path length traversed.\\

Very recently, the E866 nuclear-dependence data have been analysed
by taking into consideration the shadowing effect and comparing with 
the three different expressions (Eqs.~\ref{eq:gavin}~-~\ref{eq:baier})
for energy loss~\cite{maxim99}. Upper limits of 
$\kappa_2 < 0.75$ GeV$^2$ and $\kappa_3 < 0.10$ GeV$^2$ have been obtained.
The $\kappa_2$ limit corresponds to a constant energy loss rate
of $< 0.44$ GeV/fm, while the $\kappa_3$ limit implies 
$\Delta E < 0.046$ GeV/fm$^2 \times L^2$, where $L$ is the quark propagation
length through the nucleus. This is very close to the lower value given
by Baier et al.~\cite{bdmps} for cold nuclear matter.\\

\begin{figure}
\center
\hspace*{0.5in}
\psfig{figure=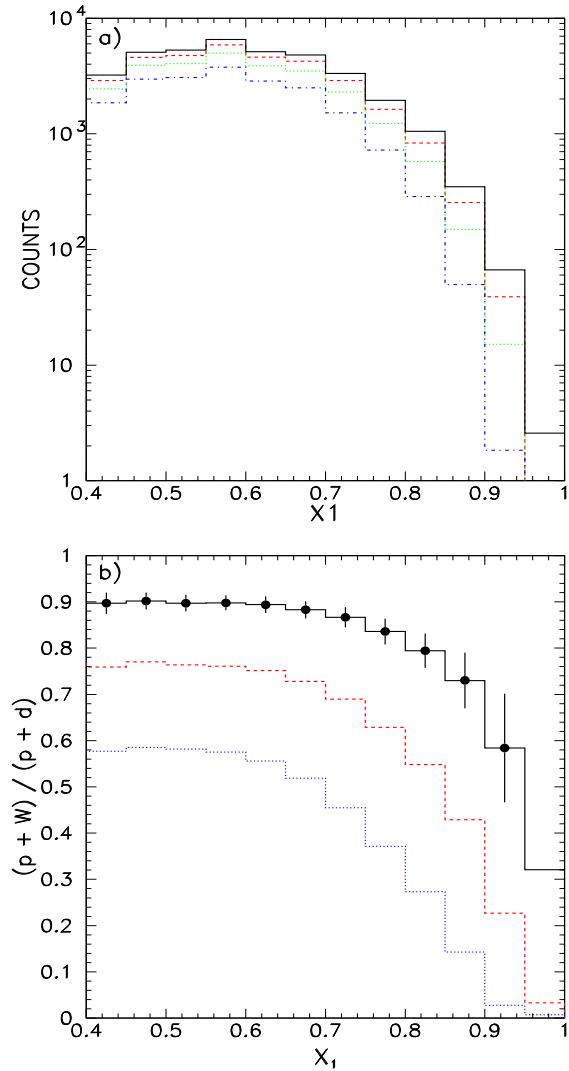,height=7.5in}
\caption{a): Solid curve is the expected $p+d$ spectrum for a 60-day run
at 50 GeV. The dashed, dotted, and dash-dotted curves correspond 
to $p+W$ spectra assuming a partonic energy loss rate of 0.1, 0.25, 0.5
GeV/fm, respectively. b): Solid circles show the expected 
statistical errors for $(p+W) / (p+d)$ ratios 
in a 60-day run for $p+W$ and $p+p$ each. The solid, dashed, and dotted
curves correspond to a partonic energy loss rate of 0.1, 0.25, 0.5
GeV/fm, respectively.}
\label{fig:elossb}
\end{figure}

A much more sensitive study of the partonic energy loss could be carried
out at the 50-GeV PS. We have simulated the effect of initial-state
energy loss on the $p+W$ Drell-Yan cross sections, and the results are 
shown in Figure~\ref{fig:elossb}. Assuming a 60-day run with the nominal
spectrometer configuration (see Section 3), the expected 
$x_1$ distribution for $p+d$
is shown as the solid curve. The dashed, dotted, and dash-dotted curves
in Figure~\ref{fig:elossb} correspond to $p+W$ $x_1$ spectra assuming
a partonic energy loss form of Eq.~\ref{eq:brodsky} with $d E / d z$
of -0.1, -0.25, -0.5 GeV/fm, respectively. The ratios of $p+W$ over $p+d$, 
shown in Figure~\ref{fig:elossb}, are very sensitive to the quark energy
loss rate, and the expected statistical accuracy can easily identify
an energy loss as small as 0.1 GeV/fm. The greater sensitivity at 50 GeV
is due to the $1/s$ factor in Eq.~\ref{eq:brodsky} and Eq.~\ref{eq:baier}.
Another important advantage at 50 GeV is the absence of shadowing effect,
and no shadowing correction is required.\\

\begin{figure}
\center
\hspace*{0.5in}
\psfig{figure=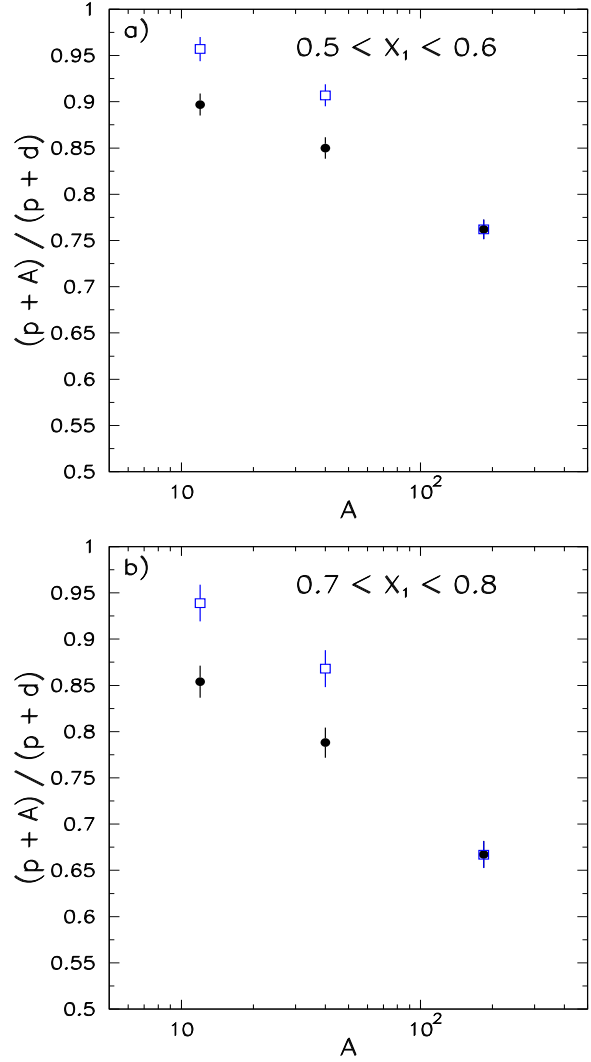,height=7.5in}
\caption{a): Solid circles correspond to the expected $(p+A) / (p+d)$
ratios assuming a partonic energy loss rate of 0.25 GeV/fm with a
nuclear dependence given by Eq.~\ref{eq:brodsky}. The open squares correspond
to partonic energy loss given by Eq.~\ref{eq:baier}. The statistical erorrs
were calculated assuming a 60-day run for each target. b) Same as the top
figure, but for a different $x_1$ bin ($0.7 < x < 0.8$).}
\label{fig:elossa}
\end{figure}

The Drell-Yan A-dependence data could further be used to determine whether the
energy loss follows an $L$ (as in Eq.~\ref{eq:brodsky}) or an 
$L^2$ (as in Eq.~\ref{eq:baier}) dependence. This is illustrated in 
Figure~\ref{fig:elossa}, where the solid circles correspond to $(p+A) / (p+d)$
assuming an energy-loss rate of 0.25 GeV/fm using Eq.~\ref{eq:brodsky}. The 
open squares correspond to the situation when energy loss is described by
Eq.~\ref{eq:baier} (the value of $\kappa_3$ is selected by matching the
$(p+W) / (p+d)$ values for both cases). Figure~\ref{fig:elossa} shows that
one can easily distinguish an $L$- from an $L^2$-dependence even when the
energy loss rate is as small as 0.25 GeV/fm.
 
\subsection{Quarkonium Production at 50 GeV}

Unlike the Drell-Yan process, the mechanisms for $J/\Psi$ production
are not well understood. Several quarkonium production models have been 
considered in the literature, including color-evaporation, color-singlet,
and color-octet models. For simplicity, we consider the color-evaporation
model, which is capable of describing the energy-dependence and the shape
of the differential cross sections well. However, the absolute normalization
of this model is treated as a parameter.\\

\begin{figure}
\center
\hspace*{0.5in}
\psfig{figure=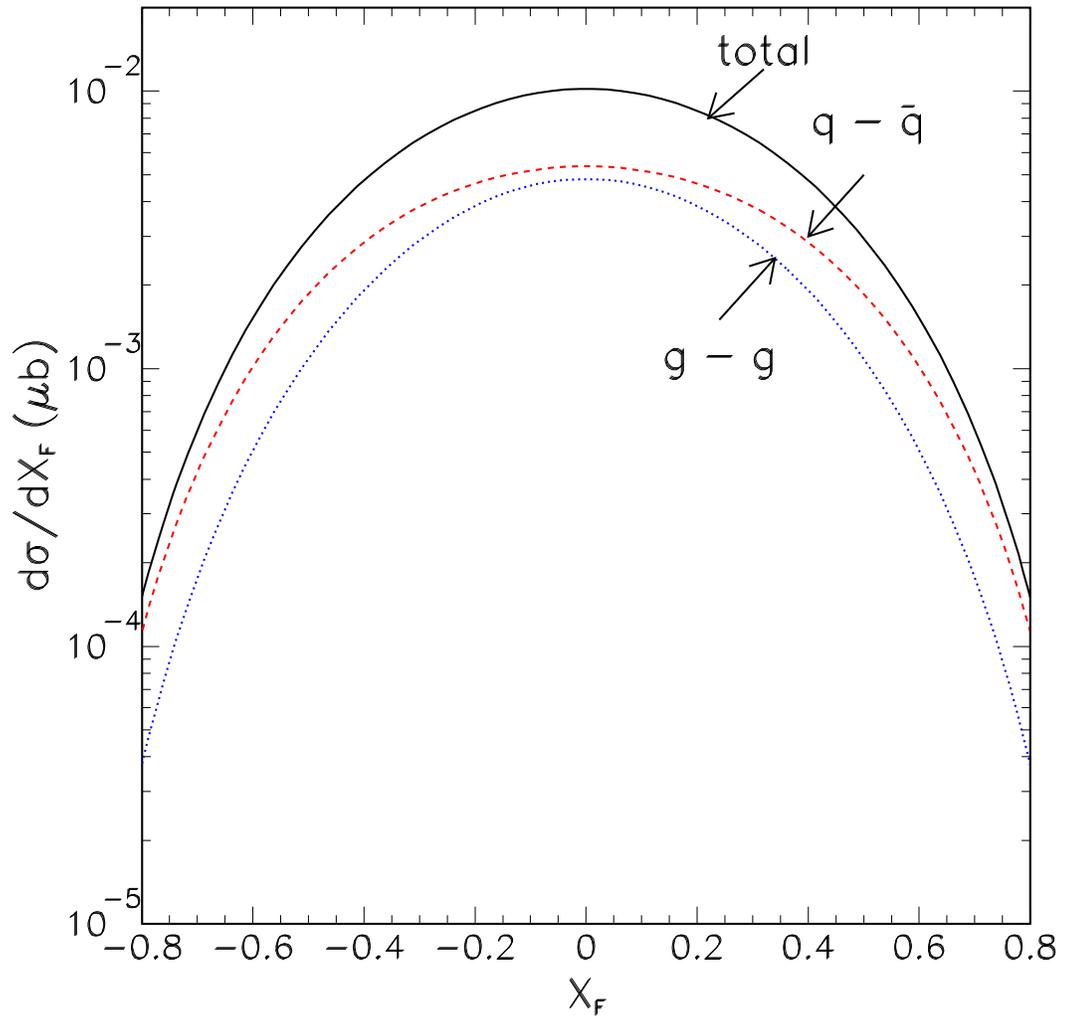,height=6.0in}
\caption{Calculation of the $p + d \to J/\Psi + x$ cross sections
at 50 GeV using the color-evaporation model. The contributions from
the gluon-gluon fusion and the quark-antiquark annihilation subprocesses
are also shown.}
\label{fig:jpsi301}
\end{figure}

Figure~\ref{fig:jpsi301} shows the prediction of the color-evaporation 
model for $J/\Psi$ production at 50 GeV. The absolute normalization is
obtained from an extrapolation of the global fit of existing $J/\Psi$ 
data~\cite{e789jpsi}. Unlike the situation at 800 GeV where the gluon-gluon
fusion subprocess dominates~\cite{e789matt}, Figure~\ref{fig:jpsi301}
shows that the quark-antiquark annihilation is the dominant subprocess
at 50 GeV. While this is reminescent of the Drell-Yan process, it is worth 
noting that quarkonium production is a hadronic process unlike the 
electromagnetic Drell-Yan process. Hence, there is no $e_q^2$ weighting 
factor for the $q - \bar q$ subprocess in $J/\Psi$ production.\\

\begin{figure}
\center
\hspace*{0.5in}
\psfig{figure=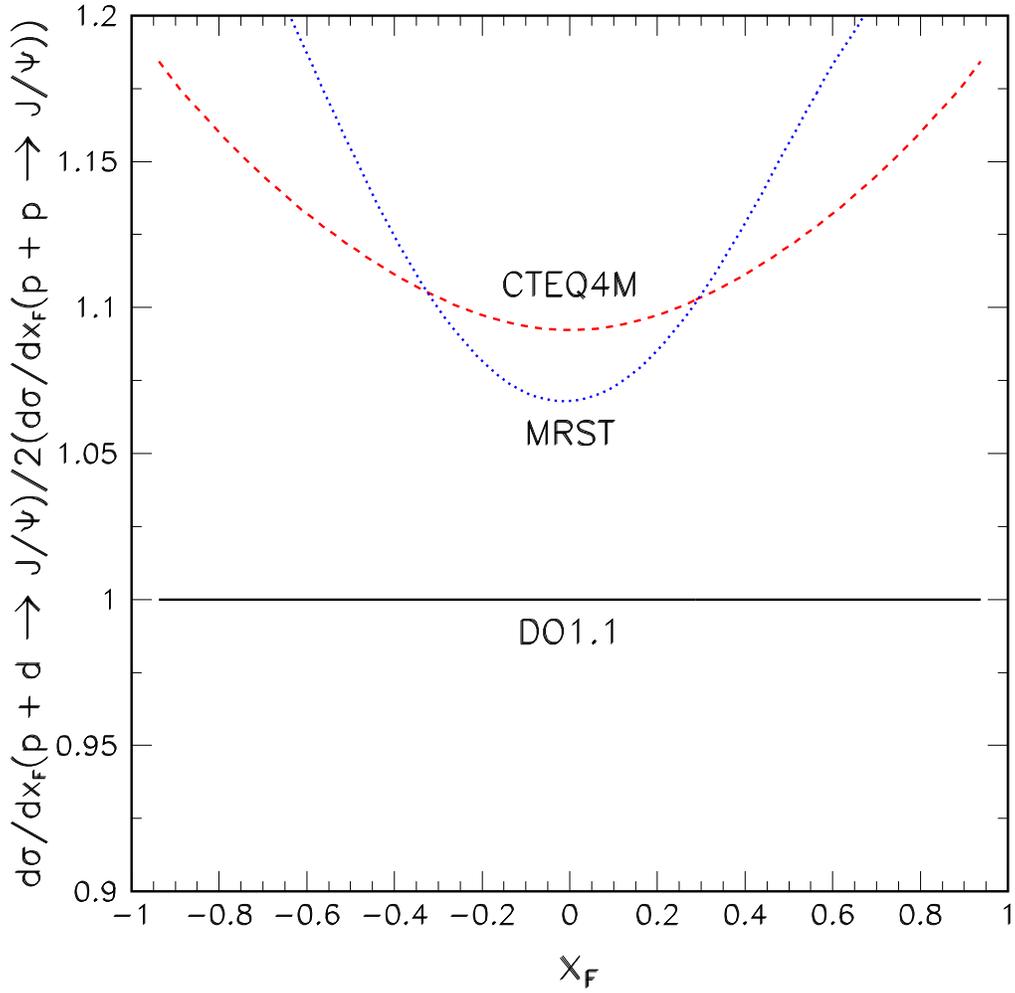,height=6.0in}
\caption{Calculations of the $p + d \to J/\Psi$ over $p + p \to J/\Psi$
ratios at 50 GeV using the color-evaporation model. The 
$\bar d / \bar u$-symmetric structure functions DO1.1 and the
$\bar d / \bar u$ asymmetric structure functions (MRST and CTEQ4M)
have been used in these calculations.}
\label{fig:jpsi310}
\end{figure}

As indicated in Figure~\ref{fig:jpsi301}, the $J/\Psi$ production data at
50 GeV are largely sensitive to quark distributions and could provide
information similar to Drell-Yan. This is illustrated in 
Figure~\ref{fig:jpsi310} which shows that the $J/\Psi$ cross
section ratio for $p + d$ over $p + p$ is very sensitive to 
the $\bar d / \bar u$ asymmetry just like the Drell-Yan process.
This could be readily tested at the 50-GeV PS, since the $J/\Psi$ event
rate is expected to be very high.\\

It is also of interest to measure $J/\Psi$ production using polarized proton
beam and target. Unfortunately, the uncertainty of the production mechanism
might make it difficult to deduce information on polarized structure
functions.

\section{Experimental Apparatus}

\begin{figure}                
\vspace*{1.5in}
\centerline{
\mbox{\rotate[r]{\epsfysize=5.5in\epsffile{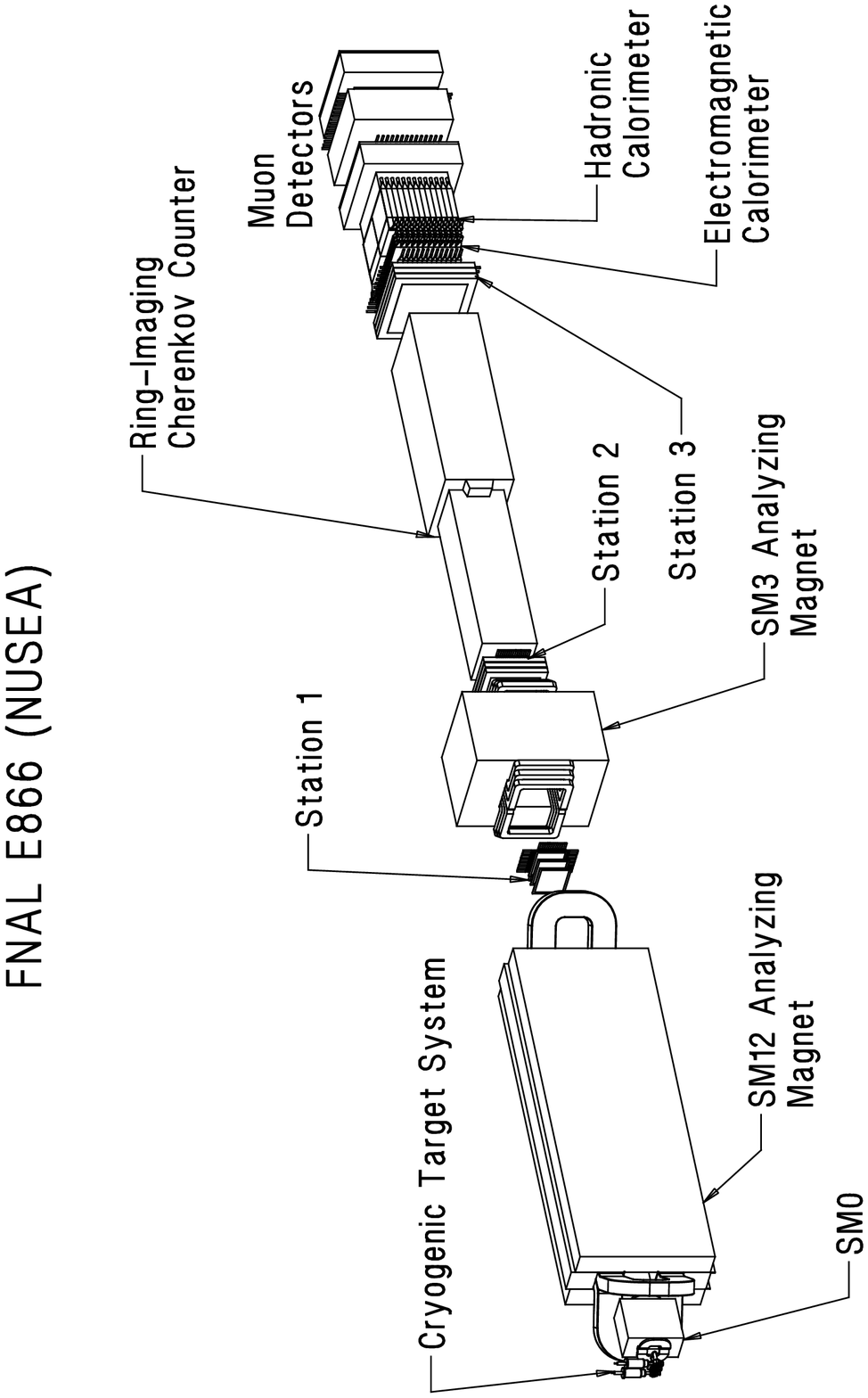}}}}
\caption{ Schematic layout of the Meson-East focusing
spectrometer at Fermilab.}
\label{fig:e866setup}         
\end{figure}

The spectrometer considered here is designed to measure 
muon pairs at $M_{\mu^{+}\mu^{-}} \geq$ 1 GeV with 50 GeV proton beam.  
The E866 spectrometer and its daughter, a proposed 
P906 spectrometer \cite{p906}, are taken as a starting point.  
The E866 spectrometer is shown in Figure~\ref{fig:e866setup}.  
$\mu^{+}\mu^{-}$ pairs produced at the target were analyzed by a 
vertical-bending spectrometer.  The remaining proton 
beam was intercepted by a copper beam dump located inside the dipole 
magnet.  The beam dump was followed by a set of absorbers made of copper, 
carbon, and polyethylene, which absorbed many of the pions and kaons
produced at the target before they could decay into muon backgrounds.
Trigger hodoscopes, muon identifiers, and tracking counters
followed the magnets. The magnetic current can be adjusted to
optimize the acceptance of a selected mass range. 
The spectrometer has good acceptance 
for dimuons with $x_F > 0$ and $p_T$ up to 3 GeV/c.\\

To design a spectrometer suitable for 50 GeV proton beam, it is
useful to consider some kinematics of the Drell-Yan process.
Table \ref{tab:kinem} compares the total center-of-mass energy 
and the Lorentz factor for proton beams of 50 GeV (at the 
present project), 120 GeV (at Fermilab Main Injector) and 
800 GeV (at Fermilab Tevatron).

\begin{table}[htbp]
  \centering
  \caption{The center-of-mass energy and the Lorentz factor for three 
           beam energies}
  \label{tab:kinem}
  \begin{tabular}{|l|l|l|l|}
    \hline
                 & 50 GeV    & 120 GeV    & 800 GeV \\
    \hline 
    $\sqrt{s}$   & 9.865 GeV & 15.12 GeV  & 38.79 GeV \\
    $\gamma_{f}$ & 5.259     & 7.998      & 20.65 \\
    \hline
  \end{tabular}
\end{table}
\noindent The Lorentz factor of the nucleon-nucleon center-of-mass frame is
\begin{equation}
\gamma_{f} = \frac{E_{1} + m_{2}}{\sqrt{s}}.\
\end{equation}
For $\mu^{+}$ and $\mu^{-}$ emitted at 90$^{\circ}$ in the 
nucleon-nucleon center-of-mass frame and for $x_F \approx 0$, 
the opening angle $\theta$ 
of the two muons in the laboratory frame is expressed as
\begin{equation}
  \tan{(\theta / 2)} = \frac{1}{\gamma_{f}}.
  \label{eq:theta}
\end{equation}
For muons emitted at 90$^\circ$ in the virtual-photon rest frame, 
the laboratory kinematics of the muons largely depends on $M$ and $x_2$ 
and does not depend on beam energy (or $\sqrt{s}$).  More specifically, 
\begin{equation}
p_{\bot}^{lab}  =  M/2.
\end{equation}
\begin{equation}
p_{\|}^{lab} = \frac{1}{2} p_{\|}^{\gamma, lab}
\simeq  \frac{1}{2} x_{1} p^{beam} 
\simeq \frac{x_{1} s}{4m_{N}} = \frac{M^{2}}{2x_{2}m_{N}}.
\end{equation}

The measurable $p_{t}$ range of muons should be almost the same 
as the E866 spectrometer or slightly smaller because the mass 
($M_{\mu^{+}\mu^{-}}$) range interested is the same or slightly 
less.  Thus the total magnetic rigidity ($\int B dl$) of the 
magnets should be about 7 to 10 T$\cdot$m, though it depends on 
the geometrical layout of the detectors.  The P906 spectrometer 
has a magnetic rigidity of 8 T$\cdot$m.\\

According to Eq.~\ref{eq:theta}, the opening angle of 
the muons at 50 GeV is about 4 times larger than at 800 GeV (E866), 
and 1.5 times larger than at 120 GeV (P906).  
One idea to design the 50-GeV spectrometer is just to shorten the existing 
setup in the beam (z) direction with the factor of 
$\gamma$(50 GeV)/$\gamma$(800 GeV) or $\gamma$(50 GeV)/$\gamma$(120 GeV).
However, since the maximum field strength of the P906 magnet is 
already near the saturation point, a magnet of almost the same 
length and wider aperture need to be considered.\\  

\begin{figure}
\center
\hspace*{0.5in}
\psfig{figure=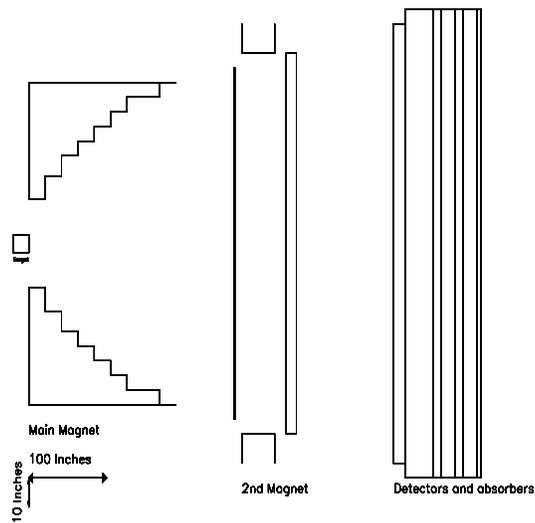,height=7.5in}
\caption{A schematic view of the prototype spectrometer.
           The top is the horizontal view and the bottom is the
           vertical view.}
\label{fig:50gevspec}
\end{figure}

Figure~\ref{fig:50gevspec} shows the horizontal and  vertical 
view of the proposed spectrometer.  The target is assumed to 
be a liquid hydrogen or 
deuterium target 20-inch long and 3-inch wide.  The produced 
charged particles are analyzed by a vertical-bending magnet, which 
is basically the same as the P906 magnet but has a wider aperture.    
The length of the magnet along the beam axis is 480.06 cm.  The 
horizontal gap of the magnet at the exit is 116.84 cm and the 
vertical gap at the exit is 279.4 cm.  The total momentum kick 
by this magnet is about 2.5 GeV/c.  
The incident proton beam is stopped by a copper beam dump, followed 
by a set of absorber materials.  The second magnet and detectors 
are placed after the first magnet.  The momentum kick by the second 
magnet is 0.5 GeV/c.  The total length of the spectrometer system 
from the entrance of the first magnet to the end of the detector system 
is 1474.47 cm.  The muons which hit all the detectors are accepted as 
signals.\\  

A fast Monte-Carlo code, which takes into account the Drell-Yan cross 
section and the spectrometer configuration, has been used to estimate 
a statistical error for $\sigma(pd)/2\sigma(pp)$ shown in Figure 
\ref{fig:rate1}.  In order to estimate the yields and statistical 
errors, the following assumptions have been applied:
\begin{itemize}
\item The beam intensity is $1 \times 10^{12}$ protons/(3 sec.).
\item The net efficiency of data acquisition is 0.5.
\item Data are taken for 60 days each for 50-cm long proton and deuteron 
      targets.
\end{itemize}
The performance of the spectrometer for $\vec p + \vec p$ measurement,
nuclear dependence study of Drell-Yan, and $J/\Psi$ production, has also
been simulated and the results have been presented in the previous
Section.

\section{Summary}

We present a broad range of physics topics which can be pursued
at the 50-GeV PS using a dimuon spectrometer and a primary proton
beam of $10^{12}$ per spill. The expected sensitivities of various
measurments have been simulated for a preliminary design of the
dimuon spectrometer. The physics scope can be considerably enlarged
with the addition of polarized proton beam and with heavy-ion beams.
More detailed studies using GEANT-based simulation are in progress
to address the issues of background and to optimize the
design of the spectrometer. Based on our study thus far, it is clear
that a rich physics program can be mounted using the primary proton
beam at the 50-GeV PS.

\newpage

\vfill
\eject

\vfill
\eject

\vfill
\eject

\vfill
\eject

\vfill
\eject

\vfill
\eject

\vfill
\eject

\vfill
\eject

\vfill
\eject

\vfill
\eject

\vfill
\eject

\vfill
\eject

\vfill
\eject

\vfill
\eject

\hspace*{-1.5in}

\end{document}